\documentclass[pra,aps,twocolumn,superscriptaddress]{revtex4}
\usepackage{hyperref}
\usepackage{bm}
\usepackage{epsfig}
\usepackage{amssymb}
\usepackage{graphicx}
\usepackage{amsbsy}
\usepackage{amsmath}

\newcommand{\kb}[2]{|#1\rangle\langle#2|}
\newcommand{\bk}[2]{\langle#1|#2\rangle}
\newcommand{\ket}[1]{|#1\rangle}

\begin{document}
\title{Nonexistence of entangled continuous-variable Werner states with positive partial transpose}

\author{Daniel McNulty}
\thanks{daniel.mcnulty@upol.cz}
\affiliation{Department of Optics, Palack\' y University,
17.~listopadu 12,  771~46 Olomouc, Czech Republic}
\author{Richard Tatham}%
\thanks{richard.tatham@dunelm.org.uk}
\affiliation{School of Physics and Astronomy, University of St.
Andrews, North Haugh, St. Andrews, Fife, KY16 9SS, Scotland}
\affiliation{Department of Optics, Palack\' y University,
17.~listopadu 12,  771~46 Olomouc, Czech Republic}
\author{Ladislav Mi\v{s}ta, Jr.}%
\thanks{mista@optics.upol.cz}
\affiliation{Department of Optics, Palack\' y University,
17.~listopadu 12,  771~46 Olomouc, Czech Republic}

\date{27 March 2014}

\begin{abstract}

We address an open question about the existence of entangled
continuous-variable (CV) Werner states with positive partial
transpose (PPT). We prove that no such state exists by showing
that all PPT CV Werner states are separable. The separability
follows by observing that these CV Werner states can be approximated by truncating the states
into a finite-dimensional convex mixture of product states. In addition, the constituents
of the product states comprise a generalized non-Gaussian measurement which gives,
rather surprisingly, a strictly tighter upper bound on quantum discord than photon counting.
These results uncover the presence of only negative partial transpose entanglement and illustrate the complexity of
more general non-classical correlations in this paradigmatic class of genuine non-Gaussian quantum states.

\end{abstract}

\maketitle

\section{Introduction}\label{sec_1}

Convex mixtures of a maximally entangled state and a maximally
mixed state of two two-level quantum systems (qubits) represent
undoubtedly the most important mixed test states in quantum
information theory. These states, commonly called Werner states
\cite{Werner_89}, combine quantum entanglement and classical noise
in a simple way that allows for the testing of quantum information
criteria, concepts and protocols in the mixed-state domain using
analytical tools. Originally developed as an example of an
entangled state which admits a local-realistic model
\cite{Werner_89}, i.e., a state which does not violate any Bell-type inequalities,
it was shown later that suitable Werner states
may exhibit hidden nonlocality \cite{Gisin_96} and that a Werner
state admitting a local-realistic model can exhibit nonlocality in the multicopy scenario \cite{Peres_96a}. In the context of
separability, Werner states have been used to demonstrate that
separability criteria based on positive partial transposition
\cite{Peres_96b} or majorization \cite{Nielsen_01} are strictly
stronger than entropic ones. Furthermore, Werner states prove to
be suitable initial states for entanglement distillation
\cite{Bennett_96}.
Interestingly, not only entangled Werner states play an important
role in quantum information; it turns out that separable Werner
states can carry non-classical correlations, known as quantum
discord \cite{Olivier_02}, which can serve as an alternative
resource for quantum technology, e.g., in quantum illumination
\cite{Weedbrook_13}.

All aforementioned applications relate to the two-qubit Werner
states. An important test state is also obtained when we extend the
Werner state to systems with an infinite-dimensional Hilbert space,
such as optical modes. A two-mode analogue of the Werner state,
the so called continuous-variable (CV) Werner state, for two modes
$A$ and $B$ is defined as \cite{Mista_02}
\begin{equation}\label{Wernerinf}
\rho_{p}=p\sigma+(1-p)\tau,\quad 0\leq p\leq 1.
\end{equation}
Here
\begin{equation}\label{sigmainf}
\sigma=\left(1-\lambda_{1}^{2}\right)\sum_{m,n=0}^{\infty}\lambda_{1}^{m+n}|m,m\rangle\langle
n,n|
\end{equation}
is the two-mode squeezed vacuum state, where
$|m,n\rangle\equiv|m\rangle_A\otimes|n\rangle_B$ with
$|k\rangle_{j}$ being the $k$-th Fock state of mode $j$, with
$\lambda_{1}=\tanh r$ and squeezing parameter $r$, and
\begin{equation}\label{tauinf}
\tau=\left(1-\lambda_{2}^{2}\right)^{2}\sum_{m,n=0}^{\infty}\lambda_{2}^{2(m+n)}|m\rangle\langle
m|\otimes|n\rangle\langle n|
\end{equation}
is the tensor product of two identical thermal states
characterized by the parameter $\lambda_{2}=\tanh s$. For $r=s$
and in the strong squeezing limit $r\rightarrow\infty$, the state
(\ref{Wernerinf}) represents a direct analogy of the original
two-qubit Werner state by approaching a convex mixture of a
maximally entangled state and a maximally mixed state in an
infinite-dimensional Hilbert space.

The CV Werner state (\ref{Wernerinf}) is a simple mixed
non-Gaussian state and therefore proves to be an excellent tool
for investigating many concepts in quantum information in the
mixed-state non-Gaussian scenario. This involves analyses of
separability, teleportation and violation of discrete-variable
Bell inequalities \cite{Mista_02}, as well as for non-classical
correlations beyond entanglement \cite{Tatham_12}. Besides, CV
Werner states have been studied also from the point of view of
violation of the CV Bell inequalities \cite{Borges_12},
quantification of non-Gaussian entanglement by negativity
\cite{Lund_08,Rodo_08} or optimality of Gaussian attacks in CV
quantum key distribution \cite{Lund_08}. Despite considerable
progress in understanding many aspects of the CV Werner states,
their basic separability properties are still not fully known.
Analysis of their separability has been performed using the
positive partial transposition (PPT) criterion \cite{Peres_96b}.
The criterion says that any two-mode \emph{separable} density
matrix $\rho$ has a positive partial transpose $\rho^{T_{A}}$,
which is a matrix with entries
\begin{equation}\label{PT}
\langle m, \mu|\rho^{T_A}|n,\nu\rangle\equiv\langle n,
\mu|\rho|m,\nu\rangle.
\end{equation}
From the PPT criterion it then follows that a quantum state is entangled if its density
matrix has a negative partial transpose (NPT). The CV Werner state (\ref{Wernerinf})
is exceptional because its NPT region can be found analytically \cite{Mista_02}.
However, this region may not contain all entangled CV Werner
states since PPT entangled states may also exist \cite{PHorodecki_97}.
In Ref.~\cite{Mista_02} a set of all PPT CV Werner states and a nontrivial proper subset
of separable states have been found. Therefore, there still exist PPT CV Werner states
for which the separability properties are not known. In particular, it is unknown whether
PPT entangled CV Werner states exist.

In the subsequent sections we answer this existence question in
the negative. First, we prove that any PPT $N$-dimensional
truncation of the CV Werner state (\ref{Wernerinf}) is separable
for any finite $N$. We then show that any PPT CV Werner state can
be approximated in the trace-norm by a sequence of its truncated
separable counterparts, which implies its separability. The
separability of the PPT truncated $N\times N$ CV Werner states is
demonstrated by finding explicitly their decomposition into a
convex mixture of pure product states, which is inspired by the
method \cite{Kraus_00} developed for simpler $2\times N$ quantum
systems. Contrary to intuition, the projectors onto the
constituents of the product states comprise a generalized
non-Gaussian measurement which yields, for a particular example of
the partial transpose of a specific PPT CV Werner state, a
strictly tighter upper bound on quantum discord than photon
counting.

The first result of the present paper closes a long-standing open
problem about the existence of PPT CV entangled Werner states. The
second result shows that more sophisticated non-Gaussian
measurements are needed to optimally extract non-classical
correlations from non-Gaussian quantum states on
infinite-dimensional Hilbert state spaces.

The paper is organized as follows. In Section~\ref{sec_2} we prove
the separability of all PPT finite-dimensional truncations of CV
Werner states, while Section~\ref{sec_3} is dedicated to the proof
of the separability of any PPT CV Werner state. In Section
\ref{discord} we show that photon counting does not minimize
quantum discord for a certain family of partially transposed CV
Werner states. Finally, Section~\ref{sec_4} contains conclusions.

\section{Finite dimensions}\label{sec_2}

At the outset we will investigate the separability of PPT states
obtained by truncating the CV Werner states (\ref{Wernerinf}) onto
a finite-dimensional Hilbert space. For two optical modes $A$ and
$B$ the truncation is defined as
\begin{equation}\label{Werner}
\rho_{p,N}=\mathcal{N}_{p,N}\left[p\sigma_{N}+(1-p)\tau_{N}\right],\quad 0\leq p\leq 1.
\end{equation}
Here,
\begin{equation}\label{rhoNOPA}
\sigma_{N}=\left(1-\lambda_{1}^{2}\right)\sum_{m,n=0}^{N-1}\lambda_{1}^{m+n}|m,m\rangle\langle n,n|
\end{equation}
is the truncated two-mode squeezed vacuum state, and
\begin{equation}\label{rhoT}
\tau_{N}=\left(1-\lambda_{2}^{2}\right)^{2}\sum_{m,n=0}^{N-1}\lambda_{2}^{2(m+n)}|m\rangle\langle
m|\otimes|n\rangle\langle n|
\end{equation}
is the tensor product of two identical truncated thermal states.
The normalization factor is given by
\begin{equation}\label{calNpN}
\mathcal{N}_{p,N}=\left[p\left(1-\lambda_{1}^{2N}\right)+\left(1-p\right)\left(1-\lambda_{2}^{2N}\right)^2\right]^{-1}.
\end{equation}

There are two PPT regions for the state $\rho_{p,N}$ which can
be distinguished depending on the relation between parameters
$\lambda_{1}$ and $\lambda_{2}$ \cite{Mista_02}.

i) For $\lambda_{1}>\lambda_{2}^{2}$ the state (\ref{Werner}) is
PPT if and only if
\begin{equation}\label{pN}
p\leq
\frac{1}{1+\frac{(1-\lambda_{1}^{2})}{(1-\lambda_{2}^{2})^{2}}\left(\frac{\lambda_{1}}{\lambda_{2}^{2}}\right)^{2N-3}}\equiv
p_{N}.
\end{equation}

ii) For $\lambda_{1}\leq\lambda_{2}^{2}$ the state (\ref{Werner})
is PPT if and only if
\begin{equation}\label{p2}
p\leq p_{N=2}\equiv p_{2}.
\end{equation}
Note, that while region (ii) coincides with the region of
PPT CV Werner states of infinite dimension \cite{Mista_02}, region (i) varies from the infinite case,
but it approaches the region of infinite-dimensional PPT CV Werner states characterized by the
condition $p=0$ \cite{Mista_02} in the limit of $N\rightarrow\infty$.

Let us start with an analysis of separability of the simple PPT
boundary state ($\equiv \rho_{q,N}$) from region (ii) for which
$\lambda_1=\lambda_{2}^{2}$ and
\begin{equation}\label{p0}
p=p_{2}=\frac{1-\lambda_{1}}{2}\equiv q.
\end{equation}
According to the definition \cite{Werner_89}, a density matrix
$\rho$ is separable if it can be written or approximated in the
trace norm by the states of the form
\begin{equation}\label{separable}
\rho=\sum_{i}p_{i}\rho_{A}^{(i)}\otimes\rho_{B}^{(i)},\quad 0\leq p_{i}\leq 1,\quad \sum_{i}p_{i}=1,
\end{equation}
where $\rho_{j}^{(i)}$ is a local density matrix of mode $j$. In order to
investigate the separability of $\rho_{q,N}$ it is easier to analyze the simpler
partially transposed state $\rho_{q,N}^{T_{A}}$, since the separability
of one implies the separability of the other. From Eqs.~(\ref{PT}) and (\ref{Werner})
we get for the partially transposed state the expression
\begin{eqnarray}\label{normWernerp0TAN}
\rho_{q,N}^{T_{A}}&=&\mathcal{K}_{N}\sum_{m,n=0}^{N-1}\lambda^{m+n}
\left(|n,m\rangle\langle m,n|+|m,n\rangle\langle m,n|\right),\nonumber\\
\end{eqnarray}
where we set $\lambda_1=\lambda$ and
\begin{eqnarray}\label{calNN}
\mathcal{K}_{N}=\frac{\left(1-\lambda^{2}\right)\left(1-\lambda\right)}{2\left(1-\lambda^{N}\right)\left(1-\lambda^{N+1}\right)}
\end{eqnarray}
is the normalization factor. Here, and in what follows, we will
sometimes use the unnormalized state
$\tilde{\rho}_{q,N}^{T_{A}}=\rho_{q,N}^{T_{A}}/\mathcal{K}_{N}$
for brevity.

Let us first start our separability analysis with the simplest
case $N=2$. Here, PPT implies separability and thus the state
$\tilde{\rho}_{q,2}^{T_A}$
must be separable. Making use of a method developed in
\cite{Kraus_00} we can find explicitly a decomposition of the
state into a convex mixture of pure product states.

The construction given in \cite{Kraus_00} involves finding and
subtracting product vectors from a state $\rho$ on the space
$\mathbb{C}^2\otimes\mathbb{C}^N$ such that $\rho$ and
$\rho^{T_A}$ remain positive. If enough product vectors exist so
that $\rho$ reduces to zero then the state is obviously separable.
The first step in finding such a decomposition for $N=2$ is
to calculate
$\tilde{r}=\text{rank}(\rho)+\text{rank}(\rho^{T_A})$. If
$\tilde{r}\leq 3N =6$ then finding the product vectors involve
calculating the roots of a polynomial. If, however, $\tilde{r}>6$
then one can always subtract product vectors from $\rho$ to reduce
its rank such that $\tilde{r}\leq6$ without affecting positivity.
In particular, one must find vectors $\ket{e,f}$ in the range of
$\rho$ such that $\ket{{e^*},f}$ lies in the range of
$\rho^{T_A}$, where $\ket{{e^*}}$ denotes complex conjugation of
$\ket{e}$. Once $\tilde{r}\leq 6$ we must then solve a polynomial
to find the remaining product states in the decomposition.

For the specific state $\tilde{\rho}_{q,2}$ we first subtract
$\ket{0,0}+\lambda\ket{1,1}$ to reduce its rank by one, such that
$\tilde{r}=6$. It then remains to find the roots $\alpha$ of the
polynomial $\text{det}[M(\alpha,{\alpha^*})]$ for the $(2\times
2)$ matrix $M$ given by
\begin{equation}
M(\alpha,{\alpha^*})=\left( \begin{array}{c}
\alpha\bk{\psi_1}{0}+\bk{\psi_1}{1} \\
{\alpha^*}\bk{\psi_2}{0}+\bk{\psi_2}{1} \end{array} \right),
\end{equation}
where $\ket{\psi_1}$ and $\ket{\psi_2}$ are the $(4\times 1)$ basis vectors for the
one-dimensional kernels of $\tilde{\rho}_{q,2}$
and $\tilde{\rho}^{T_A}_{q,2}$, respectively, and
$\langle\psi_{j}|i\rangle=\sum_{l=0}^{1}\psi_{il}^{(j)\ast}\langle
l|$ with $\langle i,l|\psi_j\rangle=\psi_{il}^{(j)}$. We then
arrive at the polynomial equation
$\alpha{\alpha^*}\sqrt{\lambda}-1=0$ which we solve to find four
more product vectors of the form
\begin{equation}\label{prodvec}
\ket{e,f}=(\alpha\ket{0}_A+\ket{1}_A)\otimes \sum_{k=0}^{1} f_k\ket{k}_B.
\end{equation}
Here the vector $\ket{f}$ is found by calculating the kernel of
matrix $M$, i.e., $M(\alpha,{\alpha^*})\vec{f}=0$ with $\vec{f}=(f_0,f_1)^{T}$, where $f_{i}=\langle i|f\rangle$, $i=0,1$. After
rewriting in a more compact form, the original state can be
expressed as $\tilde{\rho}_{q,2}^{T_{A}}=D^{(2)}$, where
\begin{eqnarray}\label{D}
D^{(2)}&=&\sum_{n=0}^{1}\lambda^{2n}\kb{n,n}{n,n}+\frac{1}{4}\sum_{n=0}^{3}\kb{q_n,q_n}{q_n,q_n},\nonumber\\
\end{eqnarray}
and $\ket{q_n}=\ket{0}+e^{i\frac{\pi n}{2}}\sqrt{\lambda}\ket{1}$.

Moving now to the case of $N>2$, we cannot directly apply the
previous method as it has been developed for only $2\times N$
systems \cite{Kraus_00}. Nevertheless, the structure of the
decomposition for $N=2$ can still inspire us to find a similar
separable decomposition for the state
$\tilde{\rho}_{q,N}^{T_{A}}$. In particular, we will show that
$\tilde{\rho}_{q,N}^{T_{A}}=D$, for $N>2$, where
\begin{multline}\label{DN}
D=\sum_{n=0}^{N-1}\lambda^{2n}\kb{n,n}{n,n}\\+\frac{1}{N^{N-1}}\sum_{n_1,\ldots,n_{N-1}=0}^{N-1}\kb{q_{\vec
n},q_{\vec n}}{q_{\vec n},q_{\vec n}},
\end{multline}
with $\vec n=(n_1,\ldots, n_{N-1})$ and $\ket{q_{\vec
n}}=\ket{0}+\sum_{j=1}^{N-1} e^{i\frac{2\pi}{N}n_j}\lambda^{j/2}\ket{j}$.

We find the matrix elements $D_{jk,lm}=\langle j,k|D|l,m\rangle$
of the decomposition (\ref{DN}) as
\begin{eqnarray}\label{DelementsN}
D_{jk,lm}&=&\lambda^{2j}\delta_{kj}\delta_{lj}\delta_{mj}+\frac{1}{N^{N-1}}(\sqrt{\lambda})^{j+k+l+m}\nonumber\\&&\times\sum_{n_1,\ldots,
n_{N-1}=0}^{N-1}e^{i\frac{2\pi}{N}(n_j+n_k-n_l-n_m)},
\end{eqnarray}
where $n_0\equiv0$. Making use of the simple relation
\begin{equation}\label{orthogonality}
\sum_{n_1,\ldots, n_{N-1}=0}^{N-1}e^{i\frac{2\pi}{N}(n_j-n_k)}=N^{N-1}\delta_{jk}
\end{equation}
to find
\begin{multline}\label{orthog}
\sum_{n_1,\ldots,
n_{N-1}=0}^{N-1}e^{i\frac{2\pi}{N}(n_j+n_k-n_l-n_m)}=\\N^{N-1}\left(\delta_{jl}\delta_{km}+\delta_{jm}\delta_{kl}-\delta_{jk}\delta_{jl}\delta_{jm}\right),
\end{multline}
we arrive at
\begin{equation}\label{Delements}
D_{jk,lm}=\lambda^{j+k}(\delta_{jl}\delta_{km}+\delta_{jm}\delta_{kl}).
\end{equation}
Note that Eq. (\ref{orthog}) is only valid for $N>2$. Since
(\ref{Delements}) are just the matrix elements
$(\rho_{q,N}^{T_A}/\mathcal{K}_{N})_{jk,lm}$, we have shown that
the state (\ref{normWernerp0TAN}) really can be expressed as a
convex mixture of product states (\ref{separable}), which reads
explicitly as
\begin{equation}\label{rhoqNdecTA}
\rho_{q,N}^{T_A}=\mathcal{K}_{N}D,
\end{equation}
where the operator $D$ is given in Eq.~(\ref{DN}), and therefore the state
is separable. Consequently, the original truncated CV Werner state
$\rho_{q,N}$ can be expressed as the following convex mixture of
product states
\begin{equation}\label{rhoqNdec}
\rho_{q,N}=\mathcal{K}_{N}D^{T_A},
\end{equation}
and therefore it is also separable.

We will now investigate the separability of the state $\rho_{p,N}$
for all other values of $p$ and show that it is in fact separable
when the inequalities (\ref{pN}) and (\ref{p2}) are satisfied. In
other words, for any PPT region the truncated CV Werner state is
separable. Our previous derivation of the separable decomposition
for $\rho_{q,N}^{T_{A}}$ will prove useful to show this.

Firstly we rewrite the state (\ref{Werner}) as
\begin{multline}\label{Wernerpdecomposition}
\rho_{p,N}=\alpha\sum_{m,n=0}^{N-1}\lambda_1^{m+n}\left(|m,m\rangle\langle n,n|+|m,n\rangle\langle m,n|\right)\\
+\sum_{m,n=0}^{N-1}\left[\beta\lambda_2^{2(m+n)}-\alpha\lambda_1^{m+n}\right]|m,n\rangle\langle
m,n|,
\end{multline}
with $\alpha=p(1-\lambda_1^2)\mathcal{N}_{p,N}$ and
$\beta=(1-p)(1-\lambda_2^2)^2\mathcal{N}_{p,N}$. Now the first sum
of (\ref{Wernerpdecomposition}) is equal to
$\tilde{\rho}_{q,N}\equiv\rho_{q,N}/\mathcal{K}_{N}$ and can be
replaced by $D^{T_A}$ from (\ref{DN}) with $\lambda=\lambda_1$.
After further rewriting it follows that the state is separable if
\begin{equation}\label{alphabetaineqality}
\sum_{m\neq
n=0}^{N-1}\left[\beta\lambda_2^{2(m+n)}-\alpha\lambda_1^{m+n}\right]|m,n\rangle\langle
m,n|\geq 0,
\end{equation}
which is equivalent to the condition
$\beta\lambda_2^{2(m+n)}-\alpha\lambda_1^{m+n}\geq0$ for $m\neq
n$. This is identical to the condition
\begin{equation}\label{pconditionmn}
p\leq\frac{1}{1+\frac{(1-\lambda_1^2)}{(1-\lambda_2^2)^2}\left(\frac{\lambda_1}{\lambda_2^2}\right)^{m+n}}.
\end{equation}
For the region $\lambda_1\leq\lambda_2^2$, the RHS of
(\ref{pconditionmn}) is minimal when $m+n=1$ and therefore the
state is separable when
\begin{equation}\label{pcondition2}
p\leq
p_2=\frac{1}{1+\frac{(1-\lambda_1^2)}{(1-\lambda_2^2)^2}\frac{\lambda_1}{\lambda_2^2}},
\end{equation}
which, according to Eq.~(\ref{p2}), is the same condition for
positive partial transposition. In particular, on the boundary
$\lambda_1=\lambda_2^2$, the inequality simplifies to $p\leq
q=\frac{1-\lambda_1}{2}$.
Finally, for the case $\lambda_1>\lambda_2^2$, the RHS of
(\ref{pconditionmn}) is minimized when $m+n$ is maximal. This
occurs when $m+n=2N-3$ and hence we arrive at inequality
(\ref{pN}) given by
\begin{equation}\label{pconditionN}
p\leq\frac{1}{1+\frac{(1-\lambda_1^2)}{(1-\lambda_2^2)^2}\left(\frac{\lambda_1}{\lambda_2^2}\right)^{2N-3}}.
\end{equation}
Since this region of $p$ accommodates all PPT states for
$\lambda_1>\lambda_2^2$, there is no room left for PPT entangled states.

\section{Infinite dimensions}\label{sec_3}
As we have just seen, for finite $N$ the truncated CV Werner state
$\rho_{p,N}$ is never entangled when its partial transposition is
positive. In fact, the same statement holds also for the
infinite-dimensional CV Werner state $\rho_{p}$ in Eq.~(\ref{Wernerinf}),
as we will show now.

Similarly to Eq.~(\ref{Wernerpdecomposition}), we can
decompose the state $\rho_{p}$ as
\begin{eqnarray}\label{Wernerpinfdecomposition}
\rho_{p}&=&\frac{\alpha'}{\mathcal{J}}\bar{\rho}_{q}+\sum_{m\ne n=0}^{\infty}\left[\beta'\lambda_2^{2(m+n)}-\alpha'\lambda_1^{m+n}\right]|m,n\rangle\langle
m,n|\nonumber\\
&&+\beta'\sum_{n=0}^{\infty}\lambda_2^{4n}|n,n\rangle\langle
n,n|,
\end{eqnarray}
where $\mathcal{J}=(1-\lambda_1)^2$, $\alpha'=p(1-\lambda_{1}^{2})$, $\beta'=(1-p)(1-\lambda_{2}^{2})^2$, and
\begin{eqnarray}\label{rhoqprimed}
\bar{\rho}_{q}&=&\mathcal{J}\left(\frac{\rho_{q}}{\mathcal{K}}-\sum_{n=0}^{\infty}\lambda_{1}^{2n}
|n,n\rangle\langle n,n|\right),
\end{eqnarray}
where
\begin{eqnarray}\label{rhoq}
\rho_{q}&=&\mathcal{K}\sum_{m,n=0}^{\infty}\lambda_{1}^{m+n}
\left(|m,m\rangle\langle n,n|+|m,n\rangle\langle m,n|\right)\nonumber\\
\end{eqnarray}
and $\mathcal{K}=(1-\lambda_{1}^2)(1-\lambda_{1})/2$ is the
normalization factor. Obviously, the state $\rho_{p}$ is separable
if both the density matrix $\bar{\rho}_{q}$ as well as the first sum on the
right-hand side of Eq.~(\ref{Wernerpinfdecomposition}) are separable quantum
states. As in the finite-dimensional case, the sum describes a
separable quantum state if the inequality (\ref{pconditionmn}) is
fulfilled. Hence, for $\lambda_1\leq\lambda_2^2$, the sum is a
separable state if the inequality (\ref{pcondition2}) is satisfied, while for
$\lambda_1>\lambda_2^2$ the inequality (\ref{pconditionN}) is
replaced by $p=0$. These regions all agree with the PPT regions
for the CV Werner state and therefore, for all PPT CV Werner states,
the first sum on the right-hand side of Eq.~(\ref{Wernerpinfdecomposition}) is always a
separable quantum state. Consequently, if the density matrix
(\ref{rhoqprimed}) is a separable state for all $\lambda_{1}$, then all
CV Werner states (\ref{Wernerinf}) with a positive partial
transposition are separable.

Unfortunately, in the limit of $N\rightarrow\infty$, the product
decomposition (\ref{rhoqNdec}) for the state $\rho _{q,N}$ does
not generalize straightforwardly and thus the separability of the
density matrix (\ref{rhoqprimed}) is not obvious. However, when
the dimension of the Hilbert space is infinite, we can use the
limit definition of separability \cite{Werner_89,Clifton_99} to
prove the separability of the state $\bar{\rho} _{q}$, given in
Eq.~(\ref{rhoqprimed}). According to the definition in
\cite{Eisert_02}, a density matrix $\rho$ is separable if there
exists a sequence $\{\rho_{n}\}_{n=1}^{\infty}$ of density
matrices $\rho_{n}$ such that each $\rho_{n}$ can be expressed as
a convex mixture of product states (\ref{separable}), and such
that
\begin{equation}\label{sep_cond_inf}
\lim_{n\rightarrow\infty}\|\rho-\rho_{n}\|_1=0,
\end{equation}
where $\|.\|_{1}$ is the trace norm.

It is again convenient to prove the separability of the partially
transposed state $\bar{\rho}_{q}^{T_A}$. The candidate for the sequence of
separable states approximating the state
$\bar{\rho}_{q}^{T_A}$ in trace norm is the sequence of states
\begin{eqnarray}\label{rhoqprimedTAN}
\bar{\rho}_{q,N}^{T_{A}}&=&\mathcal{J}_{N}\left(\frac{\rho_{q,N}^{T_{A}}}{\mathcal{K}_{N}}-\sum_{n=0}^{N-1}\lambda_{1}^{2n}
|n,n\rangle\langle n,n|\right),
\end{eqnarray}
where $\mathcal{J}_{N}=[(1-\lambda_{1})/(1-\lambda_{1}^{N})]^{2}$,
and $\mathcal{K}_{N}$ and $\rho_{q,N}^{T_A}$ are defined in
Eqs.~(\ref{calNN}) and (\ref{rhoqNdecTA}), respectively, where
$\lambda$ is replaced with $\lambda_{1}$. In other words, the
sequence $\{\rho_{n}\}_{n=1}^{\infty}$ in our case reads
$\rho_{n}\equiv\bar{\rho}_{q,n+2}^{T_A}$, $n=1,2,\ldots$.
Therefore, our goal is to verify the limit (\ref{sep_cond_inf})
for the Hermitian operator $\bar{\rho}_{q}^{T_A}-\rho_{n}$ with
$\rho_{n}=\bar{\rho}_{q,N}^{T_A}$, where here, and in what
follows, we use the label $N=n+2$ for simplicity. The trace norm
of a Hermitian operator $X$ is defined as
$\|X\|_{1}\equiv\mbox{Tr}\sqrt{X^{\dag}X}$ \cite{Reed_72} and it
is equal to the sum of the absolute values of the eigenvalues of
the operator $X$ \cite{Vidal_02}. The operator
$\bar{\rho}_{q}^{T_A}-\bar{\rho}_{q,N}^{T_A}$ has a block-diagonal
form with $1\times1$ blocks in the one-dimensional subspaces
$\mathcal{H}^{(l)}$ spanned by the vectors $\{|l,l\rangle\}$,
$l=0,1,2,\ldots$ and $2\times2$ blocks in the two-dimensional
subspaces $\mathcal{H}^{(j,k)}$ spanned by the vectors
$\{|j,k\rangle,|k,j\rangle\,j>k\}$, $j,k=0,1,2,\ldots$. On each
subspace $\mathcal{H}^{(j,k)}$ the operator
$\bar{\rho}_{q}^{T_A}-\bar{\rho}_{q,N}^{T_A}$ has one zero
eigenvalue and one nonzero eigenvalue of the form
\begin{equation}\label{ejk}
e^{(jk)}=\left\{
    \begin{array}{ll}
      2(\mathcal{J}-\mathcal{J}_{N})\lambda^{j+k} , & j,k=0,1,\ldots,N-1 \\
     2\mathcal{J}\lambda^{j+k}\,, & \hbox{otherwise}
    \end{array}
  \right.
  \end{equation}
while on the subspace $\mathcal{H}^{(l)}$ it has one eigenvalue equal to
\begin{equation}\label{el}
e^{(l)}=\left\{
    \begin{array}{ll}
      (\mathcal{J}-\mathcal{J}_{N})\lambda^{2l} , & 0\leq l\leq N-1 \\
     \mathcal{J}\lambda^{2l}\, , & N\leq l
    \end{array}
  \right.
  \end{equation}
where here and in what follows we set $\lambda_{1}=\lambda$ for brevity.
Hence, the sought trace norm reads as
\begin{widetext}
\begin{eqnarray}\label{rhoT-rhoTN1}
\|\bar{\rho}_{q}^{T_A}-\bar{\rho}_{{q},N}^{T_A}\|_{1}&=&(\mathcal{J}_{N}-\mathcal{J})\left(\sum_{k=0}^{N-1}\lambda^{2k}+2\sum_{j>k=0}^{N-1}\lambda^{j+k}\right)
+\mathcal{J}\left(\sum_{k=N}^{\infty}\lambda^{2k}+2\sum_{j=N}^{\infty}\sum_{k=0}^{N-1}\lambda^{j+k}
+2\sum_{j>k=N}^{\infty}\lambda^{j+k}\right),
\end{eqnarray}
\end{widetext}
where $\mathcal{J}_N$ is defined below Eq.~(\ref{rhoqprimedTAN}), and where we have used the inequality $\mathcal{J}\leq\mathcal{J}_{N}$.
The geometric series on the right-hand side of the latter equation can be summed, and after some algebra becomes
\begin{eqnarray}\label{rhoT-rhoTN2}
\|\bar{\rho}_{q}^{T_A}-\bar{\rho}_{q,N}^{T_A}\|_{1}&=&4\lambda^{N}-2\lambda^{2N}.
\end{eqnarray}
Returning back to the original label $n$ by substituting
$N=n+2$ we get
\begin{eqnarray}\label{rhoT-rhon}
\|\bar{\rho}_{q}^{T_A}-\rho_{n}\|_{1}&=&4\lambda^{n+2}-2\lambda^{2(n+2)}.
\end{eqnarray}
As $0\leq\lambda<1$, the right-hand side forms the $n$th element of a
convergent series and therefore in the limit $n\rightarrow\infty$ it
vanishes, i.e.,
\begin{equation}
\lim_{n\rightarrow\infty}\|\bar{\rho}_{q}^{T_A}-\rho_{n}\|_1=0.
\end{equation}
The density matrix $\bar{\rho}_{q}^{T_{A}}$ is therefore separable
and it can be expressed as a continuous convex mixture of product
states \cite{Holevo_05}. Linearity of partial transposition
preserves the structure of the state and hence also the original
density matrix $\bar{\rho}_{q}$ is separable. Consequently, all
PPT CV Werner states are separable as we set out to prove.


\section{Photon counting does not minimize discord}\label{discord}

The simplicity of CV Werner states makes them a perfect test bed
for the challenging analysis of correlations in CV non-Gaussian
states. Besides entanglement, CV Werner states can also carry a
more general form of non-classical correlations \cite{Tatham_12},
which may be present even if the state is separable. The
correlations manifest themselves through a nonzero quantum discord
\cite{Olivier_02}, which is an optimized difference of two
quantized classically equivalent expressions for mutual
information. Quantum discord is equipped with an
information-theoretical interpretation in the context of quantum
state merging \cite{Cavalcanti_11,Madhok_11}, it quantifies the
advantage of coherent quantum operations over local ones
\cite{Gu_12} and can be applied for the certification of
entangling capability of quantum gates \cite{Almeida_13}.

For a generic quantum state $\rho_{AB}$, quantum discord can be
expressed as $\mathcal{D}(\rho_{AB})=\inf_{\{\Pi_b\}}\mathcal{D}(\rho_{AB}|\left\{\Pi_{b}\right\})$,
where
\begin{equation}\label{Discord}
\mathcal{D}(\rho_{AB}|\left\{\Pi_{b}\right\})=\mathcal{S}(\rho_{B})-\mathcal{S}(\rho_{AB})+
\sum_{b}p_{b}\mathcal{S}(\rho_{A|b})
\end{equation}
is the so called measurement-dependent discord \cite{Modi_12}.
Here $\mathcal{S}(\rho_{B})$ and $\mathcal{S}(\rho_{AB})$ are von
Neumann entropies of the local state
$\rho_{B}=\mbox{Tr}_{A}(\rho_{AB})$ and the global state
$\rho_{AB}$, respectively. The state $\rho_{A|b}=\mbox{Tr}_{B}(\rho_{AB}\Pi_{b})/p_{b}$ is the
conditional state of the subsystem $A$ after the measurement
$\Pi_{b}$ on subsystem $B$ with outcome $b$, and
$p_{b}=\mbox{Tr}(\rho_{AB}\Pi_{b})$ is the probability of event $b$.

The key role in the separability analysis of PPT CV Werner states
has been played by the partial transposition
\begin{eqnarray}\label{rhoqTA}
\rho_{q}^{T_{A}}&=&\mathcal{K}\sum_{m,n=0}^{\infty}\lambda^{m+n}
\left(|n,m\rangle\langle m,n|+|m,n\rangle\langle m,n|\right)\nonumber\\
\end{eqnarray}
of the density matrix (\ref{rhoq}), which is a simple nontrivial
non-Gaussian state suitable for analysis of quantum discord. The
evaluation of the quantum discord for the state (\ref{rhoqTA})
contains a nontrivial optimization over all non-Gaussian
measurements, which is not a tractable task. For this reason we
resort to the evaluation of non-optimized discord (\ref{Discord}),
representing an upper bound on the true discord. Here we consider
two different measurements on mode $B$: firstly photon counting,
represented by the set of projectors onto Fock states
$\{\Pi_{m}=|m\rangle\langle m|\}$; and secondly the positive
operator valued measure (POVM) $\{\Pi_{\vec n},\Pi_{0}\}$. The
POVM elements $\Pi_{\vec n}$ are defined as
\begin{equation}\label{Pin}
\Pi_{\vec n}=\frac{1}{N^{N-1}}|\tilde{q}_{\vec n}\rangle\langle \tilde{q}_{\vec n}|,\quad \vec n=(n_1,\ldots, n_{N-1}),
\end{equation}
with $n_i\in\{0,\ldots,N-1\}$, and where the vectors
\begin{equation}\label{tildeqn}
|\tilde{q}_{\vec n}\rangle=\ket{0}+\sum_{j=1}^{N-1} e^{i\frac{2\pi}{N}n_j}\ket{j},
\end{equation}
are equal to the state vectors $\ket{q_{\vec
n}}=\ket{0}+\sum_{j=1}^{N-1} e^{i\frac{2\pi}{N}n_j}
\lambda^{j/2}\ket{j}$ appearing in the decomposition (\ref{DN}) of
the state $\tilde{\rho}_{q,N}^{T_{A}}$, with $\lambda=1$.  The
elements of the POVM (\ref{Pin}) are Hermitian positive-semidefinite
operators and they satisfy the completeness condition on the
$N$-dimensional space spanned by the Fock states
$\{|0\rangle,|1\rangle,\ldots,|N-1\rangle\}$, i.e.,
\begin{equation}\label{Pincompleteness}
\sum_{\vec n}\Pi_{\vec n}=\openone_{N},
\end{equation}
where $\sum_{\vec n}\equiv\sum_{n_1,\ldots,n_{N-1}=0}^{N-1}$ and
$\openone_{N}$ is the identity operator on the $N$-dimensional
space. As a consequence, if we complete the collection of
operators $\{\Pi_{\vec n}\}$ by the Hermitian
positive-semidefinite operator
\begin{equation}\label{Pi0}
\Pi_{0}=\openone-\sum_{\vec n}\Pi_{\vec n},
\end{equation}
where $\openone$ is the identity operator on a Hilbert space of a single mode,
we see that the set of operators $\{\Pi_{\vec n},\Pi_{0}\}$ comprises a single-mode POVM.

The discord (\ref{Discord}) for photon counting has been derived
in Ref.~\cite{Tatham_12} in the following simple form:
\begin{equation}\label{Dphotoncounting}
\mathcal{D}(\rho_{q}^{T_{A}}|\left\{\Pi_{m}\right\})=\lambda\ln 2.
\end{equation}
To calculate the discord for the second measurement $\{\Pi_{\vec
n},\Pi_{0}\}$, we first need to determine the global and local von
Neumann entropies ${\cal S}(\rho_{q}^{T_{A}})$ and ${\cal
S}(\rho_{q,B}^{T_{A}})$, respectively. An advantage of the
density matrix $\rho_{q}^{T_{A}}$ and its reduced density matrix
$\rho_{q,B}^{T_{A}}\equiv\mbox{Tr}_{A}[\rho_{q}^{T_{A}}]$ is that
their eigenvalues can be computed analytically, which gives the
entropies in the form \cite{Tatham_12}:
\begin{eqnarray}\label{Stilderho}
{\cal S}(\rho_{q}^{T_{A}})&=&-\left[\ln\left(2{\cal
K}\right)+ \frac{1+3\lambda}{1-\lambda^2}\lambda\ln\lambda\right]
\end{eqnarray}
and
\begin{eqnarray}\label{StilderhoB}
{\cal S}(\rho_{q,B}^{T_{A}})&=&-\left[{\cal K}\sum_{m=0}^{\infty}\left(\lambda^{2m}+\frac{\lambda^m}{1-\lambda}\right)\ln\left(\lambda^{m}+\frac{1}{1-\lambda}\right)\right.\nonumber\\
&&+\left.\ln\left({\cal
K}\right)+\frac{\lambda\left(1+3\lambda\right)}{2\left(1-\lambda^2\right)}\ln\lambda\right].
\end{eqnarray}
The remaining average entropy $\sum_{b}p_{b}\mathcal{S}(\rho_{A|b})$ from (\ref{Discord}), for the POVM $\{\Pi_{\vec n},\Pi_{0}\}$, has the structure
\begin{equation}\label{H}
\sum_{\vec n}p({\vec n})\mathcal{S}(\rho_{q,A|{\vec n}}^{T_{A}})+p_{0}\mathcal{S}(\rho_{q,A|0}^{T_{A}}).
\end{equation}
Here $\rho_{q,A|{\vec n}}^{T_{A}}$ is the conditional state of mode $A$ after
detection of the POVM element $\Pi_{\vec n}$ on mode $B$ of $\rho_q^{T_A}$, with $p({\vec n})$ the probability of outcome $\vec n$.
Similarly, $\rho_{q,A|0}^{T_{A}}$ is the conditional state of mode $A$ after detection of
the POVM element $\Pi_{0}$ on mode $B$ with probability $p_0$.

If the POVM element $\Pi_{\vec n}$ is detected on mode $B$ of state (\ref{rhoqTA}),
the unnormalized conditional state $\tilde{\rho}_{q,A|{\vec n}}^{T_{A}}
\equiv\mbox{Tr}_{B}[\rho_{q}^{T_{A}}\Pi_{\vec n}]$ of mode $A$ reads as
\begin{eqnarray}\label{tilderhon1}
\tilde{\rho}_{q,A|{\vec n}}^{T_{A}}&=&F_{\vec n}\tilde{\rho}_{A|{\vec n}}F_{\vec n}^{\dag},
\end{eqnarray}
where
\begin{eqnarray}\label{tilderhon2}
\tilde{\rho}_{A|{\vec n}}&=&\frac{\mathcal{K}}{N^{N-1}}\left(\sum_{m,n=0}^{N-1}\lambda^{m+n}|n\rangle\langle m|\right.\nonumber\\
&&\left.+\frac{1-\lambda^{N}}{1-\lambda}\sum_{m=0}^{\infty}\lambda^{m}|m\rangle\langle m|\right),
\end{eqnarray}
and
\begin{equation}\label{Fn}
F_{\vec n}=\sum_{j=0}^{\infty}e^{i\frac{2\pi}{N}n_j}|j\rangle\langle j|
\end{equation}
is the unitary operator. Making use of Eqs.~(\ref{tilderhon1})--(\ref{tilderhon2}),
and the cyclic property of the trace, we then arrive at the following expression for
the probability $p({\vec n})=\mbox{Tr}_{A}[\tilde{\rho}_{q,A|{\vec n}}^{T_{A}}]$
of detecting the measurement outcome ${\vec n}$,
\begin{equation}
p({\vec n})=\frac{\mathcal{K}}{N^{N-1}}\left[\frac{1-\lambda^{2N}}{1-\lambda^{2}}+\frac{1-\lambda^{N}}{(1-\lambda)^{2}}\right].
\end{equation}
Hence, the normalized conditional state $\rho_{q,A|{\vec n}}^{T_{A}}$ appearing in the average
entropy (\ref{H}) attains the form
\begin{eqnarray}\label{rhon1}
\rho_{q,A|{\vec n}}^{T_{A}}&=&F_{\vec n}\rho_{A|{\vec n}}F_{\vec n}^{\dag}
\end{eqnarray}
with $\rho_{A|{\vec n}}=\tilde{\rho}_{A|{\vec n}}/p({\vec n})$, where
$\tilde{\rho}_{A|{\vec n}}$ is defined in Eq.~(\ref{tilderhon2}).

If, on the other hand, the POVM element $\Pi_{0}$ is detected
on mode $B$ of state (\ref{rhoqTA}), one gets the
unnormalized conditional state $\tilde{\rho}_{q,A|0}^{T_{A}}
\equiv\mbox{Tr}_{B}[\rho_{q}^{T_{A}}\Pi_{0}]$ of mode $A$ in the form:
\begin{eqnarray}\label{tilderho0}
\tilde{\rho}_{q,A|0}^{T_{A}}&=&\mbox{Tr}_{B}[\rho_{q}^{T_{A}}(\openone-\sum_{\vec n}\Pi_{\vec n})]=
\rho_{q,A}^{T_{A}}-\sum_{\vec n}\tilde{\rho}_{q,A|{\vec n}}^{T_{A}},\nonumber\\
\end{eqnarray}
where $\rho_{q,A}^{T_{A}}\equiv\mbox{Tr}_{B}[\rho_{q}^{T_{A}}]$
is the reduced state of mode $A$ and the state $\tilde{\rho}_{q,A|{\vec n}}^{T_{A}}$
is given in Eq.~(\ref{tilderhon1}). From Eq.~(\ref{rhoqTA}) we find by direct calculation the
reduced state
\begin{equation}\label{tilderhoA}
\rho_{q,A}^{T_{A}}={\cal K}\sum_{m=0}^{\infty}\left(\lambda^{2m}+\frac{\lambda^m}{1-\lambda}\right)|m\rangle\langle m|.
\end{equation}
Using Eqs.~(\ref{tilderhon2})--(\ref{Fn}) and the orthogonality relation (\ref{orthogonality}),
we can further express the sum on the RHS of Eq.~(\ref{tilderho0}) as
\begin{eqnarray}\label{sumn}
\sum_{\vec n}\tilde{\rho}_{q,A|{\vec n}}^{T_{A}}&=&\mathcal{K}\left(\sum_{m=0}^{N-1}\lambda^{2m}|m\rangle\langle m|\right.\nonumber\\
&&\left.+\frac{1-\lambda^{N}}{1-\lambda}\sum_{m=0}^{\infty}\lambda^{m}|m\rangle\langle m|\right).
\end{eqnarray}
Substituting from Eqs.~(\ref{tilderhoA}) and (\ref{sumn}) to the RHS of
Eq.~(\ref{tilderho0}) and carrying out the trace, we arrive, after some
algebra, at the probability
\begin{equation}\label{p0}
p_{0}=\mbox{Tr}[\tilde{\rho}_{q,A|0}^{T_{A}}]=\mathcal{K}\left[\frac{\lambda^{N}}{(1-\lambda)^{2}}+\frac{\lambda^{2N}}{1-\lambda^{2}}\right]
\end{equation}
of detecting the element $\Pi_{0}$ on mode $B$ of the state (\ref{rhoqTA}).
Again making use of Eqs.~(\ref{tilderhoA}) and (\ref{sumn}) in the RHS of
Eq.~(\ref{tilderho0}), and utilizing Eq.~(\ref{p0}), we can also
derive the eigenvalues of the normalized conditional state
$\rho_{q,A|0}^{T_{A}}=\tilde{\rho}_{q,A|0}^{T_{A}}/p_{0}$ in the form:
\begin{equation}\label{fl}
f_{l}=\left\{
    \begin{array}{ll}
      \mathcal{L}_{N}\frac{\lambda^{N}}{1-\lambda}\lambda^{l}, & l=0,1,\ldots,N-1; \\
     \mathcal{L}_{N}\left(\lambda^{2l}+\frac{\lambda^{N}}{1-\lambda}\lambda^{l}\right)\, , & l=N,N+1,\ldots,
    \end{array}
  \right.
  \end{equation}
where $\mathcal{L}_{N}\equiv (p_{0}/\mathcal{K})^{-1}$.

Let us now move back to the evaluation of the average entropy (\ref{H}). The von Neumann entropy
is invariant with respect to unitary transformations and since the conditional state
$\rho_{A|{\vec n}}=\tilde{\rho}_{A|{\vec n}}/p({\vec n})$ is independent of the measurement
outcome $\vec n$, and $\sum_{\vec n}p({\vec n})=1-p_{0}$, the entropy (\ref{H}) simplifies to
\begin{equation}\label{Hsimple}
(1-p_{0})\mathcal{S}(\rho_{A|{\vec n}})+p_{0}\mathcal{S}(\rho_{q,A|0}^{T_{A}}).
\end{equation}
The second entropy $\mathcal{S}(\rho_{q,A|0}^{T_{A}})$ can be calculated from the formula
\begin{equation}\label{vNentropy}
\mathcal{S}(\rho_{q,A|0}^{T_{A}})=-\sum_{l=0}^{\infty}f_{l}\ln f_{l}
\end{equation}
where $f_{l}$ are the eigenvalues (\ref{fl}). The entropy
$\mathcal{S}(\rho_{A|{\vec n}})$ can be calculated by first
finding numerically the eigenvalues of the density matrix
$\rho_{A|{\vec n}}$ and then using Eq.~(\ref{vNentropy}). Hence,
together with Eqs.~(\ref{Stilderho}) and (\ref{StilderhoB}),
we finally get the measurement-dependent discord
(\ref{Discord})
\begin{eqnarray}\label{DPivecn}
\mathcal{D}(\rho_{q}^{T_{A}}|\left\{\Pi_{\vec n},\Pi_{0}\right\})&=&(1-p_{0})\mathcal{S}(\rho_{A|{\vec n}})+p_{0}\mathcal{S}(\rho_{q,A|0}^{T_{A}})\nonumber\\
&&+{\cal S}(\rho_{q,B}^{T_{A}})-{\cal S}(\rho_{q}^{T_{A}})
\end{eqnarray}
for the second measurement $\left\{\Pi_{\vec n},\Pi_{0}\right\}$.

\begin{figure}[tb]
\includegraphics[width=7.5cm,angle=0]{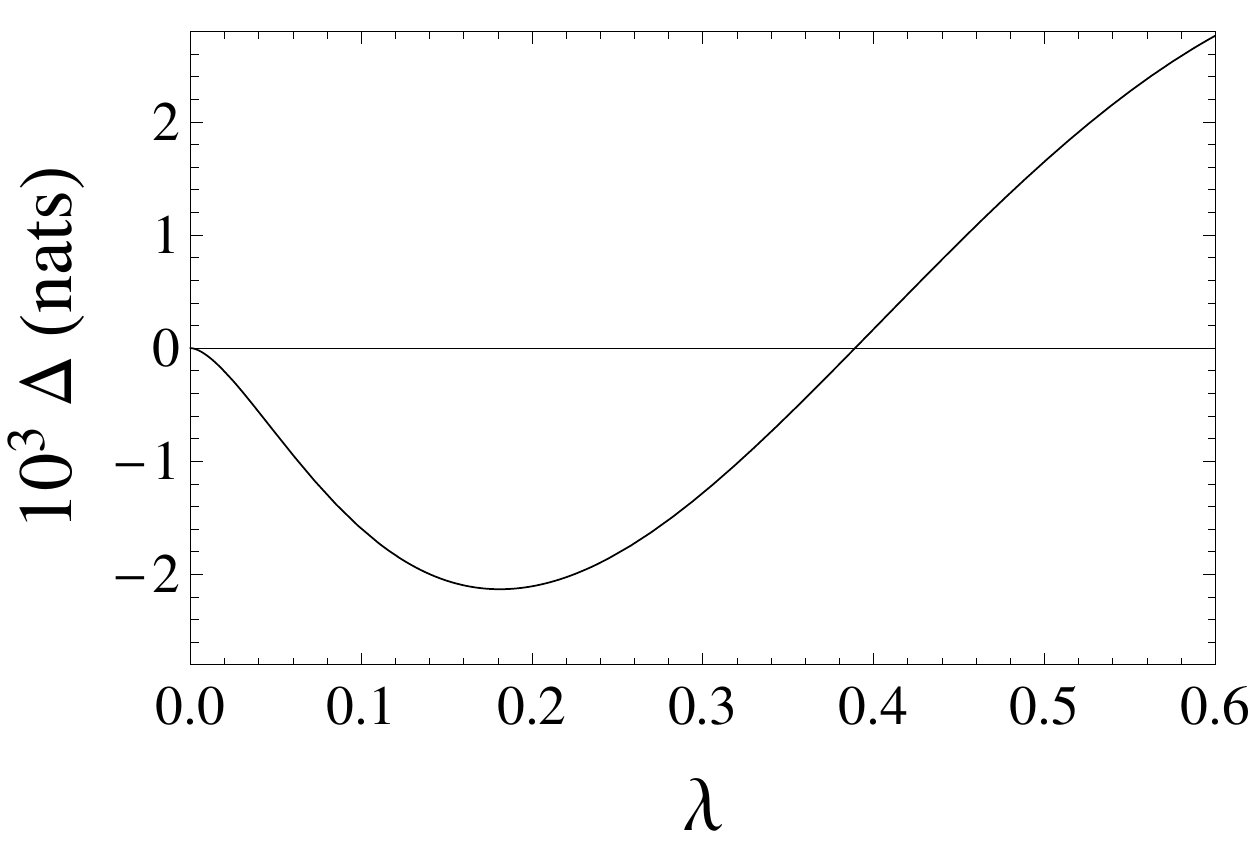}
\caption{Difference
$\Delta=\mathcal{D}(\rho_{q}^{T_{A}}|\left\{\Pi_{\vec
n},\Pi_{0}\right\})-\mathcal{D}(\rho_{q}^{T_{A}}|\left\{\Pi_{m}\right\})$
versus the parameter $\lambda$ for the state (\ref{rhoqTA}).
$\mathcal{D}(\rho_{q}^{T_{A}}|\left\{\Pi_{\vec
n},\Pi_{0}\right\})$ is the measurement-dependent discord
(\ref{DPivecn}) for measurement $\{\Pi_{\vec n},\Pi_{0}\}$ with
$N=30$ and
$\mathcal{D}(\rho_{q}^{T_{A}}|\left\{\Pi_{m}\right\})=\lambda\ln2$
is the measurement-dependent discord for photon counting
\cite{Tatham_12}. In the interval $0<\lambda\leq\lambda_{\rm
th}\doteq0.389$ the difference $\Delta$ is negative and the
measurement $\left\{\Pi_{\vec n},\Pi_{0}\right\}$ outperforms photon
counting. All infinite sums encountered in the expressions of
entropies have been approximated by $500$-th partial
sum.}\label{fig1}
\end{figure}

In Fig.~\ref{fig1} we plot the difference
$\Delta=\mathcal{D}(\rho_{q}^{T_{A}}|\left\{\Pi_{\vec
n},\Pi_{0}\right\})-\mathcal{D}(\rho_{q}^{T_{A}}|\left\{\Pi_{m}\right\})$
of measurement-dependent discords (\ref{DPivecn}) and
(\ref{Dphotoncounting}) against the parameter $\lambda$.
Inspection of the figure and further numerical analysis reveal
that in the region $0<\lambda\leq\lambda_{\rm th}\doteq0.389$ the
difference $\Delta$ is negative, thus the POVM $\left\{\Pi_{\vec
n},\Pi_{0}\right\}$ outperforms photon counting.

In Ref.~\cite{Tatham_12} it was conjectured that photon counting is
the globally optimal measurement achieving quantum discord for all
CV Werner states. The analysis of Ref.~\cite{Tatham_12} also
shows that for both the generic CV Werner states and the partially
transposed CV Werner state (\ref{rhoqTA}), upper bounds on the
most widely used quantifiers of non-classical correlations,
covering quantum discord and measurement-induced disturbance
\cite{Luo_08,Girolami_11}, coincide for photon counting.
Therefore, the properties of the partially transposed CV Werner
state (\ref{rhoqTA}) and the generic CV Werner states share
similarities as far as non-classical correlations are concerned.
In light of the fact that for a subfamily of CV Werner states
given by a convex mixture of a two-mode squeezed vacuum (\ref{sigmainf})
and vacuum, photon counting is the optimal
measurement strategy for quantum discord \cite{Tatham_12},
one may be tempted to conjecture that photon counting is also an
optimal measurement achieving quantum discord for the partially
transposed state (\ref{rhoqTA}). The present analysis disproves
this conjecture by showing that for a certain region of parameter
$\lambda$, a better measurement strategy can be found which gives
a strictly lower measurement-based discord (\ref{Discord}) than
photon counting.

Let us now investigate the physical implementation of the partial
transpose (\ref{rhoqTA}) of the PPT CV Werner state (\ref{rhoq}).
The state is a complex non-Gaussian state which lives in the
entire infinite-dimensional symmetric subspace of the two-mode
Hilbert space and it is therefore challenging to prepare
experimentally with current technology. Nevertheless, we can
prepare, at lest in principle, the $N$-dimensional truncation
$\rho_{q,N}^{T_A}$ of the state for low $N$. Specifically, from
the decomposition (\ref{DN}) it follows that the truncated state
can be obtained as a convex mixture of the (unnormalized) state
$\sum_{n=0}^{N-1}\lambda^{2n}\kb{n,n}{n,n}$ and the product states
$\kb{q_{\vec n},q_{\vec n}}{q_{\vec n},q_{\vec n}}$ with different
${\vec n}$. The first state can be created by truncating the
phase-randomized two-mode squeezed vacuum state (\ref{sigmainf})
using quantum scissors \cite{Pegg_98,Villas-Boas_01}, while the
states $|q_{\vec n}\rangle$ can be prepared conditionally using
displacements, squeezers and photon subtraction using the method
of Ref.~\cite{Fiurasek_05}.

It is also of interest to look at the physical interpretation of
the partially transposed PPT CV Werner states for $r=s$ and in the
limit of infinite squeezing $r\rightarrow\infty$. After partial
transposition, the two-mode squeezed vacuum state (\ref{sigmainf})
transforms into an operator which, in this limit, approaches the
flip operator $V=\sum_{m,n=0}^{\infty}|n,m\rangle\langle m,n|$.
Consequently, the partially transposed PPT CV Werner state
converges to a mixture of the flip operator and a maximally mixed
state, and is therefore analogous to the finite-dimensional Werner
state \cite{Werner_89,Horodecki_01}. This behaviour should be
contrasted with the asymptotic behaviour of the original CV Werner
state (\ref{Wernerinf}), which converges to the mixture of a
maximally entangled state and a maximally mixed state in infinite
dimensions. This latter mixture is analogous to the
finite-dimensional isotropic states
\cite{Horodecki_99,Horodecki_01}, which are expressed in terms of
a convex mixture of a maximally entangled state
$\sum_{j=0}^{d-1}|j,j\rangle/\sqrt{d}$ and a maximally mixed state
$\openone_{d^2}/d^{2}$, where $\openone_{d^2}$ is the
$d^2$-dimensional identity matrix. In light of this
correspondence, one could argue that it is more appropriate to
denote the state (\ref{Wernerinf}) as the CV isotropic state and
its positive partial transposition as the CV Werner state.
However, as the isotropic and Werner states are equivalent (up to
local unitary transformations) in the two-qubit scenario, one can
reason that it is legitimate to call the state (\ref{Wernerinf}) a
CV Werner state and consider it a generalization of the two qubit
Werner state.


\section{Conclusions}\label{sec_4}

We have shown that all PPT CV Werner states are separable.
Inspired by a method in \cite{Kraus_00} designed for $2\times N$
systems we have first decomposed all PPT truncated $N\times N$ CV
Werner states into a convex mixture of product states, thus proving
their separability. Next we have shown that the truncated states
approximate, in the trace norm, their infinite-dimensional
counterparts, which implies separability of all PPT CV Werner
states. Finally, we have constructed a generalized non-Gaussian
measurement from the product states of a CV Werner state decomposition and
shown that the measurement extracts more non-classical
correlations than photon counting, as quantified by quantum discord.

The results presented in this paper reveal a similarity between CV
Werner states and the original Werner states since the PPT
condition is equivalent to separability for both
\cite{Horodecki_01}. This fact may also raise the question of
whether Werner states share a similarity also for distillability.
In particular, one may ask whether, like in the finite-dimensional
case, there are NPT CV Werner states satisfying the reduction
separability criterion \cite{Horodecki_99}, which would be the
candidates for currently hypothetical NPT non-distillable
entangled states \cite{Horodecki_09}. The answer to this question
is left for future research. Besides, as the PPT CV Werner states
considered here are also elements of a larger set of PPT states
\cite{Chruscinski_06} possessing a similar structure, the present
approach can serve as a recipe on how to analyse separability of
other states from this set. We hope that our results will inspire
further studies on separability and non-classical correlations in
PPT quantum states both in finite- and infinite-dimensional
Hilbert state spaces.
\section{Acknowledgment}

D. M. acknowledges the support of the Operational Program
Education for Competitiveness Project No. CZ.1.07/2.3.00/20.0060
co-financed by the European Social Fund and Czech Ministry of
Education. L. M. would like to acknowledge the Project No.
P205/12/0694 of GA\v{C}R.


\begin{thebibliography}{99}
\bibitem{Werner_89} R. F. Werner, Phys. Rev. A {\bf 40}, 4277
(1989).
\bibitem{Gisin_96} N. Gisin, Phys. Lett. A {\bf 210}, 151 (1996).
\bibitem{Peres_96a} A. Peres, Phys. Rev. A {\bf 54}, 2685 (1996).
\bibitem{Peres_96b} A. Peres, Phys. Rev. Lett. {\bf 77}, 1413 (1996).
\bibitem{Nielsen_01} M. A. Nielsen and J. Kempe, Phys. Rev. Lett.
{\bf 89}, 5184 (2001).
\bibitem{Bennett_96} C. H. Bennett, G. Brassard, S. Popescu, B.
Schumacher, J. A. Smolin, and W. K. Wootters, Phys. Rev. Lett.
{\bf 76}, 722 (1996).
\bibitem{Olivier_02} H. Ollivier and W. H. Zurek, Phys. Rev. Lett.
{\bf 02}, 017901 (2002).
\bibitem{Weedbrook_13} C. Weedbrook, S. Pirandola, J. Thompson, V.
Vedral, and M. Gu, e-print arXiv:1312.3332.
\bibitem{Mista_02} L. Mi\v{s}ta, Jr., R. Filip, and J. Fiur\'a\v{s}ek, Phys. Rev. A {\bf 65}, 062315 (2002).
\bibitem{Tatham_12} R. Tatham, L. Mi\v{s}ta, Jr., G. Adesso, and N.
Korolkova, Phys. Rev. A {\bf 85}, 022326 (2012).
\bibitem{Borges_12} C. V. S. Borges, A. Z. Khoury, S. Walborn, P.
H. S. Ribeiro, P. Milman, and A. Keller, Phys. Rev. A {\bf 86},
052107 (2012).
\bibitem{Lund_08} A. P. Lund, T. C. Ralph, and P. van Loock, J.
Mod. Opt. {\bf 55}, 2083 (2008).
\bibitem{Rodo_08} C. Rod\'o, G. Adesso, and A. Sanpera, Phys. Rev.
Lett. {\bf 100}, 110505 (2008).
\bibitem{PHorodecki_97} P. Horodecki, Phys. Lett. A {\bf 232}, 333 (1997).
\bibitem{Kraus_00} B. Kraus, J. I. Cirac, S. Karnas, and M. Lewenstein, Phys. Rev. A {\bf 61}, 062302 (2000).
\bibitem{Clifton_99} R. Clifton and H. Halvorson, Phys. Rev. A {\bf 61}, 012108 (1999).
\bibitem{Eisert_02} J. Eisert, C. Simon, and M. B. Plenio, J. Phys. A {\bf 35}, 3911 (2002).
\bibitem{Reed_72} M. Reed and B. Simon, {\it Methods in Modern Mathematical
Physics} (Academic Press, New York, 1972), Vol. 1.
\bibitem{Vidal_02} G. Vidal and R. F. Werner, Phys. Rev. A {\bf
65}, 032314 (2002).
\bibitem{Holevo_05} A. S. Kholevo, M. E. Shirokov, and R. F. Werner, Russ. Math. Surv. {\bf 60}, 359 (2005); arXiv:quant-ph/0504204.
\bibitem{Cavalcanti_11}
D. Cavalcanti, L. Aolita, S. Boixo, K. Modi, M. Piani, and A.
Winter, Phys. Rev. A {\bf 83}, 032324 (2011).
\bibitem{Madhok_11} V. Madhok and A. Datta, Phys. Rev. A {\bf 83}, 032323
(2011).
\bibitem{Gu_12} M. Gu, H. M. Chrzanowski, S. M. Assad, T. Symul,
K. Modi, T. C. Ralph, V. Vedral, and P. K. Lam, Nat. Phys. {\bf
8}, 671 (2012).
\bibitem{Almeida_13} M. P. Almeida, M. Gu, A. Fedrizzi, M. A.
Broome, T. C. Ralph, and A. G. White, e-print arXiv:1301.7110.
\bibitem{Modi_12} K. Modi, A. Brodutch, H. Cable, T. Paterek, and
V. Vedral, Rev. Mod. Phys. {\bf 84}, 1655 (2012).
\bibitem{Luo_08} S. Luo, Phys. Rev. A {\bf 77}, 022301 (2008).
\bibitem{Girolami_11} D. Girolami, M. Paternostro, and G. Adesso,
J. Phys. A {\bf 44}, 352002 (2011).
\bibitem{Pegg_98} D. T. Pegg, L. S. Phillips, and S. M. Barnett,
Phys. Rev. Lett. {\bf 81}, 1604 (1998).
\bibitem{Villas-Boas_01} C. J. Villas-Boas, Y. Guimar\~aes, M. H.
Y. Moussa, and B. Baseia, Phys. Rev. A {\bf 63}, 055801 (2001).
\bibitem{Fiurasek_05} J. Fiur\'{a}\v{s}ek, R. Garc\'{\i}a-Patr\'{o}n, and N. J. Cerf, Phys. Rev. A {\bf 72}, 033822 (2005).
\bibitem{Horodecki_01} P. Horodecki and R. Horodecki, Quantum Inf. and Comp. {\bf 1}, 45 (2001).
\bibitem{Horodecki_99} M. Horodecki and P. Horodecki, Phys. Rev. A {\bf 59}, 4206 (1999).
\bibitem{Horodecki_09} R. Horodecki, P. Horodecki, M. Horodecki,
and K. Horodecki, Rev. Mod. Phys. {\bf 81}, 865 (2009).
\bibitem{Chruscinski_06} D. Chru\'{s}ci\'{n}ski and A.
Kossakowski, Phys. Rev. A {\bf 74}, 022308 (2006).
\end{thebibliography}
\end{document}